\documentclass[doublecol]{epl2}
\usepackage{amssymb}
\usepackage{amsbsy}
\usepackage{amsmath}

\title{Spin transport in the XXZ model at high temperatures: Classical dynamics versus quantum $S=1/2$ autocorrelations}

\shorttitle{Spin transport in the XXZ model at high temperatures}

\author{Robin Steinigeweg\thanks{E-mail: \email{robin.steinigeweg@ijs.si}}}

\shortauthor{R. Steinigeweg}

\institute{
J. Stefan Institute - Jamova 39, SI-1000 Ljubljana, Slovenia, EU
}


\abstract{The transport of magnetization is analyzed for the classical
Heisenberg chain at and especially above the isotropic point. To this end,
the Hamiltonian equations of motion are solved numerically for initial states
realizing harmonic-like magnetization profiles of small amplitude and with
random phases. Above the isotropic point, the resulting dynamics is observed
to be diffusive in a hydrodynamic regime starting at comparatively small times
and wave lengths. In particular, hydrodynamic regime and diffusion constant
are both found to be in quantitative agreement with close-to-equilibrium
results from quantum $S=1/2$ autocorrelations at high temperatures. At the
isotropic point, the resulting dynamics turns out to be non-diffusive at the
considered times and wave lengths.}

\begin{document}

\maketitle

Transport in one-dimensional systems has been a topic of theoretical
investigations for several decades, and even nowadays there is an
ongoing and still increasing interest in understanding transport
phenomena in such systems, including their dependence on temperature
\cite{casati1984, garrido2001, li2004, sirker2009, zotos1999,
heidrichmeisner2003, benz2005, prosen2011,  karrasch2011, fabricius1998,
sirker2006, grossjohann2010, znidaric2011, michel2008, prosen2009,
steinigeweg2010, steinigeweg2011-1, steinigeweg2011-2, prelovsek2004,
mierzejewski2011, langer2009, fabricius1997, mueller1988, gerling1989,
gerling1990, dealcantarabonfim1992, boehm1993}.
Much work has been devoted to a qualitative classification of the
dynamics into either non-normal ballistic or normal diffusive behavior.
In this context the crucial mechanisms for the emergence of pure
diffusion have been addressed and non-integrability is frequently
discussed w.r.t.~its role as a basic prerequisite, see \cite{sirker2009},
for instance. Particularly, spin chain models are a central issue of
research
\cite{sirker2009, zotos1999, heidrichmeisner2003, benz2005, prosen2011,
karrasch2011, fabricius1998, sirker2006, grossjohann2010, znidaric2011,
michel2008, prosen2009, steinigeweg2010, steinigeweg2011-1, steinigeweg2011-2,
prelovsek2004, mierzejewski2011, langer2009, fabricius1997, mueller1988,
gerling1989, gerling1990, dealcantarabonfim1992, boehm1993},
with a considerable focus on transport of magnetization (spin) in the
anisotropic $S=1/2$ Heisenberg chain
\cite{sirker2009, zotos1999, heidrichmeisner2003, benz2005, prosen2011,
karrasch2011, fabricius1998, sirker2006, grossjohann2010, znidaric2011,
michel2008, prosen2009, steinigeweg2010, steinigeweg2011-1, steinigeweg2011-2,
prelovsek2004, mierzejewski2011, langer2009, fabricius1997},
which serves as a prototype model due to its integrability in
terms of the Bethe ansatz \cite{bethe1931}. The specific question
of spin transport in this model has experienced an upsurge of
interest during the last decade, not least motivated by experiments
on low-dimensional quantum magnets, where large magnetic heat
conduction \cite{sologubenko2000, hess2001, hlubek2010} and long
nuclear magnetic relaxation times \cite{thurber2001, kuehne2009}
have been observed. But still, despite much effort, spin transport
in the anisotropic $S=1/2$ Heisenberg chain is not fully understood
yet, not even for finite anisotropies $\Delta \neq 0$.

So far, many works have focused on the dynamics of magnetization in
the limit of both zero momentum and frequency in connection with the
spin Drude weight, see, e.g., \cite{zotos1999, heidrichmeisner2003,
benz2005, prosen2011, karrasch2011}. While non-zero Drude weights for
$\Delta < 1$ \cite{prosen2011} imply partially ballistic dynamics
\cite{sirker2009} (for $\Delta =0$ completely ballistic dynamics) and
prevent the occurrence of purely diffusive dynamics at zero momentum
and frequency, they do not allow for conclusions on diffusion laws at
finite momentum. Early analysis of the time-dependent correlation
function of the local spin density \cite{fabricius1998} suggested the
absence of diffusion for all anisotropies $0 \leq \Delta \leq 1$ in the
high-temperature limit $T = \infty$, see also \cite{sirker2006};
however, recent low-temperature studies at the isotropic point $\Delta = 1$,
using bosonization and transfer matrix renormalization group \cite{sirker2009}
as well as Monte Carlo \cite{grossjohann2010}, are consistent with
finite-frequency diffusion at small momentum. On the other hand,
$T = \infty$ investigations at $\Delta = 1$ \cite{znidaric2011} on the
basis of the Lindblad quantum master equation point to superdiffusive
dynamics. But diffusion is found in this approach for anisotropies
$\Delta > 1$ \cite{znidaric2011, michel2008, prosen2009} with a
high-temperature diffusion coefficient which is in quantitative
agreement with time-dependent perturbation theory for the current
autocorrelation in the limit of zero momentum \cite{steinigeweg2010,
steinigeweg2011-1}. Further analysis of the spin density
autocorrelation at $\Delta > 1$ \cite{steinigeweg2011-2} demonstrated
finite-time diffusion at small momentum also for lower temperatures
$T < \infty$; however, the spectrum of the zero-momentum current
autocorrelation might still feature low-frequency anomalies
indicative for an insulator \cite{prelovsek2004, mierzejewski2011}.

In this work we will shed light on the high-temperature dynamics
of magnetization in a different way, namely, by applying classical
mechanics. Certainly, the application of classical mechanics is
not a novel concept; however, most work has focused on correlation
functions at the isotropic point $\Delta = 1$ yet \cite{mueller1988,
gerling1989, gerling1990, dealcantarabonfim1992, boehm1993} without
providing a quantitative comparison to quantum mechanics and $S=1/2$,
which cannot be considered as a classical limit a priori, even at
high temperatures $T = \infty$. Here, we will provide such a
comparison, motivated by a possibly simple scaling of
correlation functions with $S$ \cite{steinigeweg2010},
and we will particularly unveil a convincing quantitative
agreement at $\Delta > 1$. Due to this agreement, our results suggest
that signatures of quantum $S=1/2$ diffusion at $\Delta > 1$
\cite{znidaric2011, michel2008, prosen2009, steinigeweg2010,
steinigeweg2011-1, steinigeweg2011-2} can be indeed understood as a
``classical'' phenomenon.

The present paper is structured as follows: First of all, we will give
the pertinent definitions by introducing the anisotropic Heisenberg chain
in general and in particular the corresponding Hamiltonian equations of
motion in the case of classical mechanics. Thereafter we will establish a
transport scenario by specifying a certain class of initial states with
inhomogeneous magnetization profiles of small amplitude. We will afterwards
continue by discussing the properties of diffusive dynamics and in this
context we will briefly review close-to-equilibrium results on diffusion at
high temperatures in the case of quantum mechanics and $S=1/2$. To these
results we will eventually compare our numerical findings on classical
mechanics, starting above but also proceeding to the isotropic point. We
will finally close with a summary.

In this work we investigate the anisotropic Heisenberg chain (XXZ model)
with a Hamiltonian of the form
\begin{equation}
H = J \sum_r^L (S_r^x S_{r+1}^x + S_r^y S_{r+1}^y + \Delta \, S_r^z
S_{r+1}^z) \, , \label{H}
\end{equation}
where $S_r^i$ ($i = x$, $y$, $z$) are the components of the spin
$\mathbf{S}_r$ at site $r$, $L$ denotes the number of sites, $J$ represents
the exchange coupling constant, and $\Delta$ is the anisotropy parameter.

In the case of classical mechanics (CM) the spins $\mathbf{S}_r$ are
three-dimensional vectors of unit length, $|\mathbf{S}_r| = 1$, and their
dynamics is generated by the Hamiltonian equations of motion
\cite{mueller1988, gerling1989, gerling1990, dealcantarabonfim1992,
boehm1993}
\begin{equation}
\dot{\mathbf{S}}_r = \frac{\partial H}{\partial \mathbf{S}_r} \times
\mathbf{S}_r \equiv \mathbf{J}_{r-1} - \mathbf{J}_{r+1} \label{HEM}
\end{equation}
with $\mathbf{J}_{r \pm 1} = \pm J \, \mathbf{S}_r \times (S_{r \pm 1}^x,
S_{r \pm 1}^y, \Delta \, S_{r \pm 1}^z)$ as the incoming and outgoing spin
currents. Since this set of equations is non-integrable in terms of the
Liouville-Arnold theorem \cite{arnold1978, steinigeweg2009}, exact analytical
solutions can only be given for mostly trivial initial configurations. We thus
solve the set of equations numerically using a 4th order Runge-Kutta algorithm
with a fixed time step of $\delta t \, J = 0.01$; however, the presented
results in the work at hand will not depend on this particular choice of
the time step, cf.~\cite{mueller1988, gerling1989, gerling1990,
dealcantarabonfim1992, boehm1993}.
\\
In order to realize a transport scenario we choose a certain class of initial
states with non-homogenous magnetization profiles. Concretely, these initial
states read
\begin{equation}
\mathbf{S}_r(0) = \left(
\begin{array}{c}
\cos(\alpha_r) \, \sqrt{1 - A^2 \, \cos^2(q r) \, a_r^2} \\
\sin(\alpha_r) \, \sqrt{1 - A^2 \, \cos^2(q r) \, a_r^2} \\
A \cos(q r) \, a_r \\
\end{array}
\right) \, , \label{IS}
\end{equation}
where the first term of the z-component is a cosine-shaped magnetization
profile at a single momentum $q = 2 \pi k/L$ and with the total amplitude
$A \in [0,1]$. The second term of the z-component introduces local
amplitudes $a_r \in [0,1]$ drawn at random from a uniform distribution.
This term hence adds contributions at other momenta $q' \neq q$, but the
main contribution is still at $q$. The remaining $x$- and $y$-components
essentially contain local phases $\alpha_r \in [0, 2\pi]$ drawn at random
from a uniform distribution again. We emphasize that Eq.~(\ref{IS})
defines a subclass of all possible initial states, which is particularly
evident for small total amplitudes $A \ll 1$. Later we will choose a small
$A = 1/4$ in order to model ``close-to-equilibrium'' states.
Due to the random choice of local phases $\alpha_r$, these states
correspond reasonably to high temperatures. In fact, also correlation
functions at high temperatures are obtained by sampling over a set of
completely random configurations, see \cite{dealcantarabonfim1992}, for instance.

Our aim is to analyze the decay of the $q$-mode in time, i.e., the
relaxation of $S^z_q = \sum_r^L \cos(q r) \, s^z_r$. For long chains of
length $L = 18000$, as considered throughout this work, the initial value
$S^z_q(0) \approx N \, A/4$ does not depend on the concrete realization
of the random numbers in Eq.~(\ref{IS}) any further; however, the relaxation
of $S^z_q$ still differs from one realization to another. We therefore
analyze the ensemble average $M_q = 1/N \, \sum_n^N S_q^z$ over $N = 1000$
different initial states in order to obtain the ``typical'' dynamics.
One may expect that each $S_q^z$ is close to the ensemble average $M_q$ in
the thermodynamic limit $L \rightarrow \infty$. In fact, this expectation
is supported because our results for $M_q$ at large $q$ can be reproduced
for $L \ll 18000$ if $N \gg 1000$. In other words, the shorter the chain
the more initial states are needed to reach the ensemble average.

The dynamics may be called diffusive if $M_q$ satisfies a (lattice)
diffusion equation of the form
\begin{equation}
\dot{M}_q(t) = -\tilde{q}^2 \, D \, M_q(t) \, , \quad \tilde{q}^2 =
2[1-\cos(q)] \sim q^2
\label{DE}
\end{equation}
with a diffusion constant $D$, which is independent of both time $t$ and
momentum $q$. Of course, a diffusion equation is expected to be valid
solely in a hydrodynamic regime starting at sufficiently long $t$ and
small $q$, e.g., in the limit of $t \rightarrow \infty$ and $q \rightarrow
0$. Hence, in order to investigate the full time- and momentum-dependence
of the dynamics, we reformulate Eq.~(\ref{DE}) and introduce a generalized
diffusion coefficient, which given by \cite{steinigeweg2011-2}
\begin{equation}
D_q(t) = \frac{\dot{M}_q(t)}{-\tilde{q}^2 \, M_q(t)} \, . \label{D}
\end{equation}
This quantity is in the focus of the present work. It is well defined
whenever $M_q$ takes on a non-zero value, otherwise it diverges. Such a
divergence naturally emerges after the relaxation of $M_q$ due to
``fluctuations'' around zero, e.g., as arising from numerical errors;
however, we study the dynamics until the relaxation of $M_q$ here. In
any case, a divergence of $D_q(t)$ is a clear indication of non-diffusive
behavior. It is worth to mention that Eq.~(\ref{D}) is equivalent to
writing $M_q$ in the form
\begin{equation}
\frac{M_q(t)}{M_q(0)} = \exp \! \left [-\tilde{q}^2 \int_0^t \!
\mathrm{d}t' \, D_q(t') \right] \, , \label{M}
\end{equation}
which leads to a strictly exponential decay for a diffusion constant
$D_q(t) = \mathrm{const}$.

Before presenting results for the case of CM, let us also discuss the
case of quantum mechanics (QM) in detail, to which we are going to
compare afterwards. In that case, all spins become operators, their
time arguments have to be understood with respect to the Heisenberg
picture, and initial configurations are specified by a density matrix
$\rho$, which is possibly a pure state. In fact, pure states have been
considered in numerical simulations \cite{langer2009} using the
time-dependent density matrix renormalization group, yet restricted
to zero temperature. In principle the ``typical'' dynamics can be
defined again by the ensemble average $M_q(t)= 1/N \, \sum_n^N
\mathrm{Tr} [ S_q^z(t) \, \rho ]$.  Here, however, we decide to proceed
in a different way and instead define $M_q$ as the density autocorrelation
$M_q(t) = \langle S^z_q(t) S^z_{-q}(0) \rangle$, where the angles denote
the canonical equilibrium average at the inverse temperature $\beta =
1/T$, set to $\beta = 0$. This way we are going to compare to linear
response theory \cite{mahan2000} at high temperatures, i.e., the quantity
$D_q(t)$ in Eq.~(\ref{D}) becomes a generalized high-temperature
diffusion coefficient close to equilibrium.
\\
We especially make use of the perturbation theory (PT) in
\cite{steinigeweg2010, steinigeweg2011-1} for $S=1/2$ and $\Delta > 1$,
which yields a  prediction for $D_q(t)$ in the limit of $q \rightarrow 0$.
Therein, $D_{q\rightarrow 0}(t)$ is found to take on the constant value
$D_{q\rightarrow 0} \, J = 0.88/\Delta$ at a time scale $t \, J > 3.0/
\Delta$. Moreover, this value has been brought into quantitative agreement
with diffusion constants from complementary non-equilibrium bath scenarios
\cite{michel2008, prosen2009, znidaric2011} on the basis of the Lindblad
quantum master equation. In addition, $D_q(t) \approx D_{q\rightarrow 0}(t)$
has been observed in \cite{steinigeweg2011-2} at finite momenta $q \lesssim
2 \pi/9$, at least at the treatable times $t \, J \lesssim 10$, using full
exact diagonalization (ED) for a chain of length $L=18$. In view
of these findings, the starting point of the following CM simulation
will be $\Delta > 1$ and $q = 2 \pi/9$.

\begin{figure}[tb]
\centering
\includegraphics[width=0.75\linewidth]{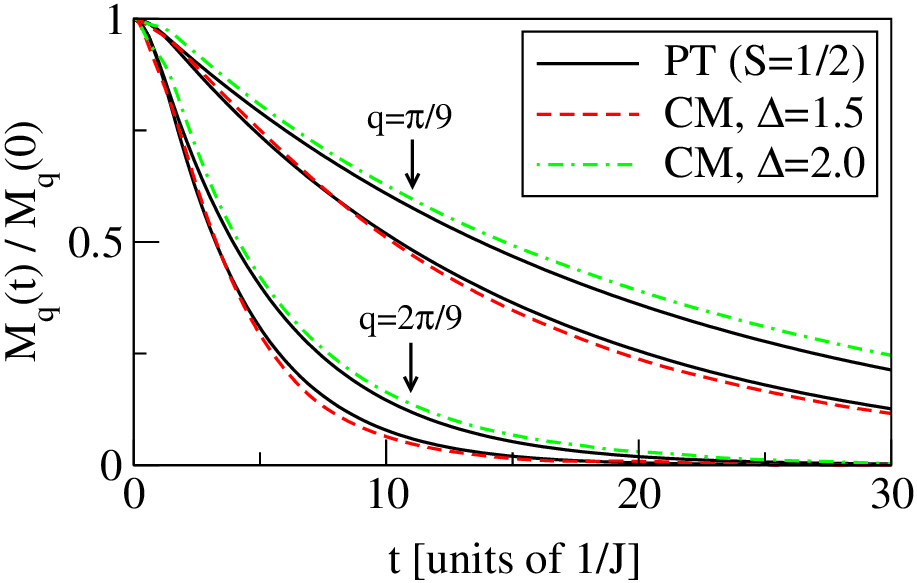}
\caption{Relaxation of modes $M_q(t)$ at momenta $q = \pi/9$, $2 \pi/9$
for anisotropies $\Delta = 1.5$, $2.0$ in a chain of length $L = 18000$,
as obtained numerically from classical mechanics for the total amplitude
$A = 1/4$ (non-solid curves). Numerical parameters: $N = 1000$, $\delta t
\, J = 0.01$. For comparison, the decay of quantum $S=1/2$ autocorrelations
at $\beta = 0$ is shown, as predicted by the $q \rightarrow 0$ perturbation
theory in \cite{steinigeweg2011-1} (solid curves).} \label{Fig1}
\end{figure}

\begin{figure}[tb]
\centering
\includegraphics[width=0.75\linewidth]{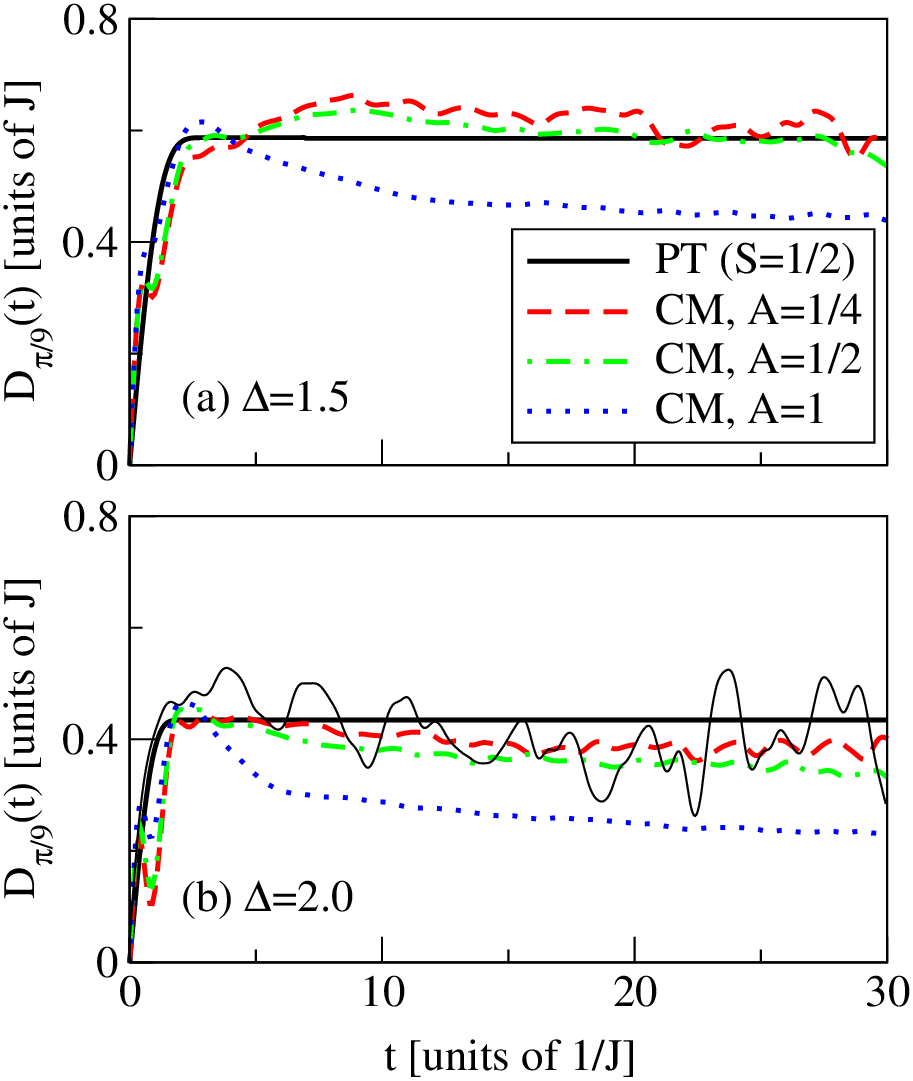}
\caption{Time-dependence of the generalized diffusion coefficient
$D_{\pi/9}(t)$ for anisotropies (a) $\Delta = 1.5$, (b) $2.0$ in a chain
of length $L = 18000$, as obtained numerically from classical mechanics
for different total amplitudes $A = 1/4$, $1/2$, and $1$ (non-solid
curves). Numerical parameters: $N = 1000$, $\delta t \, J = 0.01$.
The $q \rightarrow 0$ prediction \cite{steinigeweg2011-1} for
quantum $S = 1/2$ autocorrelations at $\beta = 0$ is indicated (thick solid
curves). In (b) the numerical $q =\pi/9$ result for the classical
{\it autocorrelation} at $\beta = 0$ is shown for $N = 10000$ (thin solid
curve).} \label{Fig2}
\end{figure}

Now we turn to our CM findings. In Fig.~\ref{Fig1} we display the
numerical result for the decay of $M_q$ for anisotropies $\Delta =1.5$,
$2.0$ at momenta $q = 2 \pi/9$, $\pi/9$, as obtained for a small total
amplitude $A = 1/4$. To repeat, we consider a chain of length $L = 18000$
and average over $N = 1000$ initial configurations. Apparently, the decay
of $M_q$ takes place at rather short time scales $t \, J \lesssim 30$.
This decay is the faster the larger $\Delta$ and the larger $q$. For
comparison, we additionally indicate the relaxation of $S = 1/2$, $\beta
= 0$ density autocorrelations, using Eq.~(\ref{M}), $D_q(t) \approx
D_{q \rightarrow 0}(t)$, as well as $D_{q \rightarrow 0}(t)$ according to
the PT in \cite{steinigeweg2011-1} (depicted in Fig.~\ref{Fig2}). The
obvious agreement in Fig.~\ref{Fig1} is remarkable for several reasons:
First, we compare the average decay of a density to the decay of a density
autocorrelation. Second, this comparison is done between CM and
QM, but generally the quantum problem can only be considered reliably as
classical for a sufficiently large spin quantum number $S$, even in the limit
of high temperatures. In any case, the agreement usually requires a proper
scaling factor taking into account the different length of a spin in CM and QM.
The latter point deserves closer attention since we have not introduced such a
scaling factor, i.e., the spins in Eq.~(\ref{HEM}) are still three-dimensional
vectors of unit length. We emphasize that this observation is not an artefact of
the applied numerical integrator because later we will reproduce existing CM
simulations at $\Delta = 1$.

Next we intend to classify the type of the dynamics in Fig.~\ref{Fig1}.
To this end, the time-dependence of the generalized diffusion coefficient
$D_q(t)$ is displayed in Fig.~\ref{Fig2} at fixed momentum $q = \pi/9$
for anisotropies $\Delta = 1.5$, $2.0$. For a small total amplitude $A=1/4$
we again find a convincing agreement with $D_{q\rightarrow0}(t)$ according
to the PT in \cite{steinigeweg2011-1}. The CM curve $D_{\pi/9}(t)$ not only
takes on an approximately constant value close to $D_{q\rightarrow0} \, J = 0.88/
\Delta$, but also reaches this value at roughly the same point in time,
namely, at $t \, J \approx 3.0/\Delta$. Clearly, the agreement is better for
$\Delta = 1.5$ in Fig.~\ref{Fig2}(a) than for $\Delta = 2.0$ in
Fig.~\ref{Fig2}(b), where the CM curve $D_{\pi/9}(t)$ slightly decreases
for $t \, J \gtrsim 10$. In order to give insight into the origin of this
decrease, we also show $D_{\pi/9}(t)$ for larger total amplitudes
$A > 1/4$. Apparently, the decrease is the more pronounced the larger $A$.
Thus, since the parameter $A$ adjusts the total amplitude of the initial
magnetization profile, we identify the decrease as a ``not-close-to-equilibrium''
effect. Even though not shown here, the decrease becomes also more pronounced
as $\Delta$ increases at fixed $A$, which is already visible in
Figs.~\ref{Fig2}(a) and (b). However, for $\Delta = 2.0$, the effect is
comparatively small for $A = 1/4$ and may completely disappear in the limit
of $A \rightarrow 0$. In our CM simulation we cannot consider the case of
a very small total amplitude $A \ll 1/4$ since $D_{\pi/9}(t)$ develops
strong ``oscillations'' if A is decreased to much. Such oscillations are
already observable for $A = 1/4$ in Fig.~\ref{Fig2} and essentially arise
from a too small number of initial states, as we will demonstrate later in
detail. This observation indicates that averaging is more important for
``close-to-equilibrium'' states, at least for a chain of finite length. It
is worth to mention that $M_q$ is less sensitive to averaging because the
integration in Eq.~(\ref{M}) smoothes the oscillatory behavior.
For completeness, we illustrate in Fig.~\ref{Fig2} (b) the
same oscillatory behavior for the classical density-density {\it correlation}
at $\beta = 0$, evaluated according to \cite{dealcantarabonfim1992} and
sampled over $N = 10000$ completely random initial states ($10$ times
more than before).  Despite the still strong oscillations, this correlation is
consistent with the decay at $A = 1/4$, which particularly excludes a
possibly ``far-from-equilibrium'' scenario for this choice of $A$.

\begin{figure}[tb]
\centering
\includegraphics[width=0.75\linewidth]{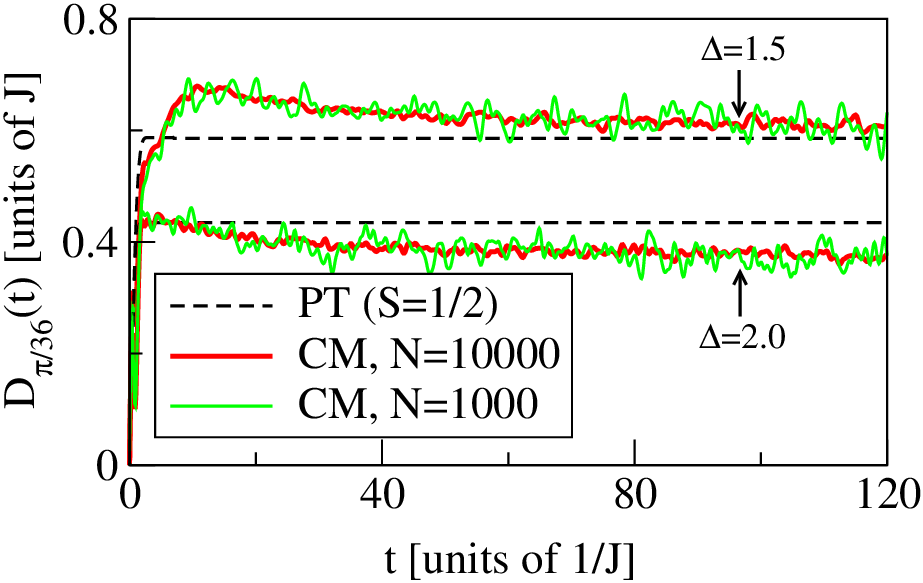}
\caption{Time-dependence of the generalized diffusion coefficient
$D_{\pi/36}(t)$ for anisotropies $\Delta = 1.5$, $2.0$ in a chain of
length $L = 18000$, as obtained numerically from classical mechanics
for the total amplitude $A = 1/4$ and different averages over $N = 1000$
and $10000$ initial states (solid curves). Remaining numerical
parameter: $\delta t \, J = 0.01$. As in Fig.~\ref{Fig2}, the quantum
$S = 1/2$, $\beta = 0$ prediction of the $q \rightarrow 0$ perturbation theory
in \cite{steinigeweg2011-1} is indicated (dashed curves).}
\label{Fig3}
\end{figure}

So far, the CM results in Figs.~\ref{Fig1}, \ref{Fig2} can be reproduced
for a chain of length $L = 18$ if the averaging is performed over $N \gg
10000$ initial configurations, which is feasible for a short chain. Hence,
these results are certainly free of relevant finite-size effects. Next we
proceed to smaller momenta $q < \pi/9$, which cannot be realized for $L = 18$
any further. In Fig.~\ref{Fig3} we display the generalized diffusion
coefficient $D_q(t)$ at fixed momentum $q=\pi/36$, still for anisotropies
$\Delta = 1.5$, $2.0$ and the small total amplitude $A=1/4$. The similarity
to the results in Fig.~\ref{Fig2} is obvious, even though a wider time window
$t \, J \leq 120$ is plotted in Fig.~\ref{Fig3}. In this time window, the
decrease of $D_{\pi/36}(t)$ appears to stagnate. On that account, without
causing a large error, we may consider $D_{\pi/36}(t)$ as an approximately
time-independent quantity,  which is not identical but at least close to
$D_{q\rightarrow0}$ from the PT in \cite{steinigeweg2011-1}. We have checked
in addition that the CM curve for $D_{q}(t)$ does not change significantly
at $q=\pi/18$ or $\pi/72$, although not shown explicitly here. This
independence of $q$, together with the approximative independence of $t$,
is a clear signature of diffusion. This observation is one main result
of the work at hand. It indicates the existence of CM diffusion
in an extended region of time and momentum in quantitative agreement
with QM predictions for $S=1/2$ at $\beta = 0$. Of course, even smaller
$q$ and longer $t$ may be evaluated; however, it is reasonable
that the above $t$- and $q$-region lies already within the hydrodynamic
regime and at some point it is hardly possible to exclude errors due to the
applied numerical integrator. In this context we demonstrate finally that
``oscillations'' of $D_q(t)$ do not refer to such errors but only depend on
the number of initial configurations. To this end, Fig.~\ref{Fig3} also
shows $D_{\pi/36}(t)$ for $N = 10000$, which clearly is smoother than
for $N = 1000$.

\begin{figure}[tb]
\centering
\includegraphics[width=0.75\linewidth]{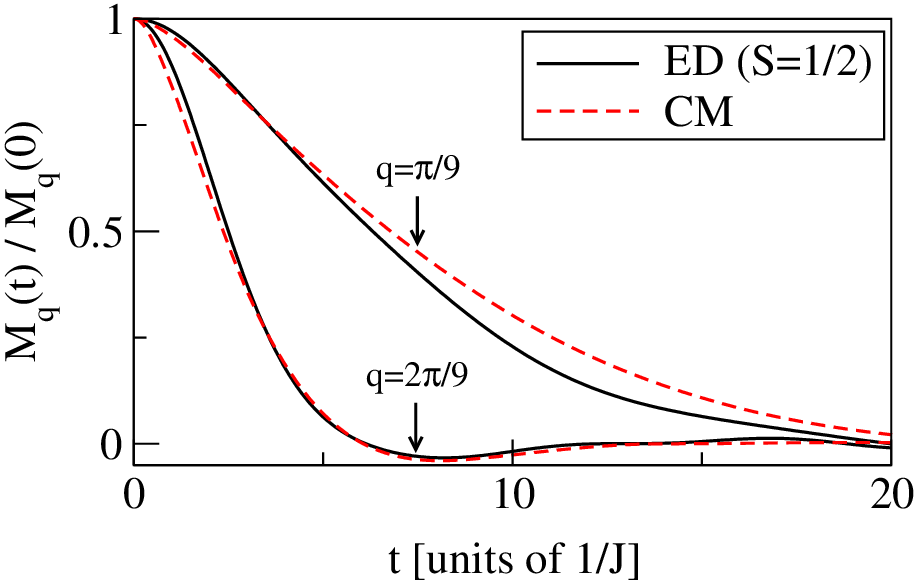}
\caption{Relaxation of modes $M_q(t)$ at momenta $q = \pi/9$, $2 \pi/9$
for the anisotropy $\Delta = 1.0$ in a chain of length $L = 18000$, as
obtained numerically from classical mechanics for the total amplitude
$A = 1/4$ (non-solid curves). Numerical parameters: $N = 1000$, $\delta t
\, J = 0.01$. For comparison, the decay of quantum $S=1/2$ autocorrelations
at $\beta = 0$ is shown, as found from exact diagonalization for a chain
of length $L = 18$.} \label{Fig4}
\end{figure}

Next we turn to the isotropic point, i.e., $\Delta = 1$. First, we intend
to clarify whether there still is a relationship to $S=1/2$, $\beta = 0$
density autocorrelations, as done before for $\Delta > 1$ in Fig.~\ref{Fig1}.
However, since the PT in \cite{steinigeweg2011-1} is not applicable in the
region $\Delta \sim 1$, we proceed differently and instead obtain these
density autocorrelations numerically using ED for a chain of length $L=18$.
To repeat, in such a short chain the smallest realized (non-zero) momentum
is $q=\pi/9$. In Fig.~\ref{Fig4} the resulting relaxation of density
autocorrelations is  displayed for momenta $q=2\pi/9$, $\pi/9$ and compared
with the CM result on the decay of $M_q$ at the same momenta but in a much
longer chain of length $L=18000$. Again, for a small total amplitude $A=1/4$
the quantitative agreement is remarkably good. Of course, the agreement
appears to be much better for the larger momentum $q=2\pi/9$, where
the same slight oscillation around zero is found in CM and QM, emerging at
a time scale $t \, J \sim 6$. Such oscillations are well known to emerge at
large $q$ \cite{fabricius1997} and zero-crossings yield divergencies of the
generalized diffusion coefficient \cite{steinigeweg2011-2}, as also visible
in later numerical data. For the smaller momentum $q=\pi/9$ the CM and QM
curves in Fig.~\ref{Fig4} start to deviate from each other at a time scale
$t \, J \sim 10$; however, finite-size effects of a QM curve for $L=18$
usually set in at this time scale \cite{steinigeweg2011-2}. On that account
the agreement in Fig.~\ref{Fig4} supports the relationship between CM and
QM, which may also hold in the limit of smaller momenta and longer times.

\begin{figure}[tb]
\centering
\includegraphics[width=0.75\linewidth]{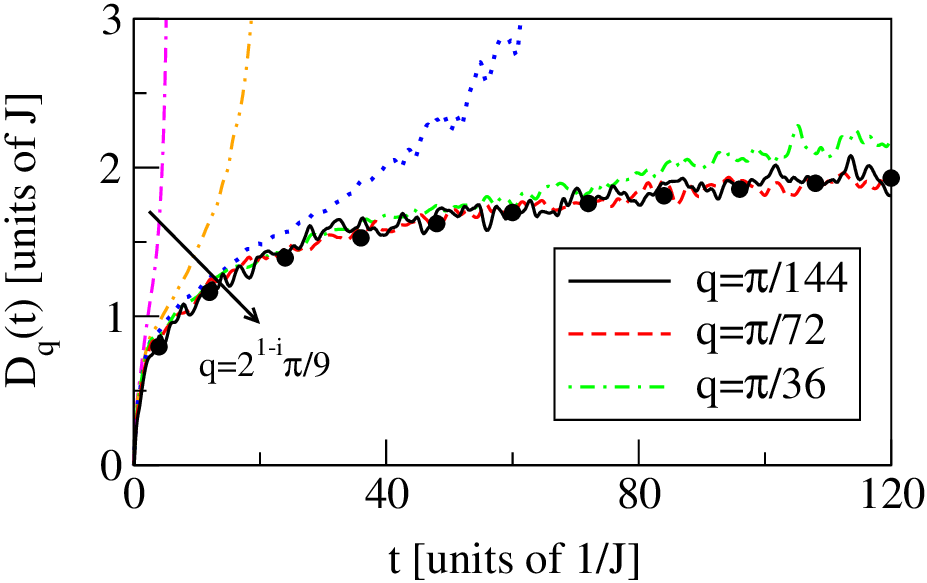}
\caption{The generalized diffusion coefficient $D_q(t)$ at momenta
$q = 2^{1-i} \, \pi/9$ ($i = 0,1,\ldots,5$) for the anisotropy $\Delta
= 1$ in a chain of length $L = 18000$, as found numerically from classical
mechanics for the total amplitude $A = 1/4$ (using $N = 1000$ initial
states and a time step of $\delta t \, J = 0.01$). For comparison,
the function $0.33 \, [1+\ln(t \, J)]$ is indicated (circles).} \label{Fig5}
\end{figure}

In Fig.~\ref{Fig5} we eventually summarize our CM results on the generalized
diffusion coefficient $D_q(t)$ for $\Delta = 1$ over a wide range of momenta
$q = 2^{1-i} \, \pi/9$ ($i=0, 1, \ldots, 5$), i.e., between $2\pi/9$ and
$\pi/144$. (Strictly speaking, the smallest momentum reads $q = 2 \pi \, 63
/18000 \approx \pi/144$.) Apparently, $D_q(t)$ depends significantly on
momentum at $q < \pi/72$ and even diverges because of the slight oscillation
of $M_q$ around zero, as already shown in Fig.~\ref{Fig4}. In the depicted
time window, $t \, J \leq 120$, a negligible dependence of $D_q(t)$ on
momentum is found firstly when varying $q$ from $\pi/72$ to $\pi/144$.
Although all CM curves in Fig.~\ref{Fig5} are consistent with $D_q(t) \sim
D_{\pi/144}(t) + \tilde{q}^4 \, (t \, J)^2/4$, this $q$-scaling has to be taken
as a rule of thumb for the dominant dependence on $q$ without derivation. 
We emphasize that the CM curve at $q = \pi/144$ agrees well with
previous results on CM density-density {\it correlations}. For instance, using in
Eq.~(\ref{D}) the function $\exp[-0.537 \, q^2 (1 + 0.1 \, \ln \! |q|) \, J \, t \, \ln(J \, t)]$,
which is one possible fit function in \cite{dealcantarabonfim1992} (see also the
discussion in \cite{boehm1993}), we obtain for $D_{\pi/144}(t)$ a corresponding
function $0.33 \, [1+\ln(t \, J)]$, as indicated in Fig.~\ref{Fig5} (cirlces). Clearly,
the time-dependence of $D_{\pi/144}(t)$ may be a pointer to such a logarithm.
Furthermore, because of the good agreement with \cite{dealcantarabonfim1992}, this
logarithmic increase cannot be understood as a ``not-close-to-equilibrium'' effect
resulting from the choice of the total amplitude $A=1/4$. In fact, we do not find a
relevant dependence on $A$. One might be tempted to speculate on the time- and
momentum-dependence of $D_q(t)$ beyond the depicted time window, see also a
related discussion in \cite{dealcantarabonfim1992,boehm1993}; however, it is
hardly feasible to verify a possibly logarithmic time-dependence of $D_q(t)$
from a merely numerical simulation. In any case, at least for the considered
$t$ and $q$ in Fig.~\ref{Fig5} the dynamics turns out to be non-diffusive. In
view of the agreement with $S=1/2$, $\beta=0$ density autocorrelations in
Fig.~\ref{Fig4}, we may expect a similar dynamics in QM. This expectation has
impact on numerical approaches to QM dynamics, simply since most approaches
are confined to the here considered $t$ and $q$, e.g., $q=\pi/144$ represents
the smallest momentum in a long chain of length $L=288$. While this length
will never be reached by ED, it may be treatable by time-dependent density
matrix renormalization group \cite{langer2009}, even though restricted to
zero temperature yet. It is worth to mention that a recent result for $L=256$
from a non-equilibrium bath scenario on the basis of the Lindblad quantum
master equation supports superdiffusive transport at $\beta = 0$
\cite{znidaric2011}, which is consistent with Fig.~\ref{Fig5}.

In summary, we studied the transport of magnetization for the classical
Heisenberg chain at and especially above the isotropic point. To this end,
we solved the Hamiltonian equations of motion numerically for initial states
realizing harmonic-like magnetization profiles of small amplitude and with
random phases. Above the isotropic point, i.e., for large anisotropies $\Delta
\sim 1.5$ we found that the resulting dynamics of the ensemble average becomes
diffusive in a hydrodynamic regime starting at rather small times and large
momenta. We further brought the diffusion constant into good quantitative
agreement with quantum $S=1/2$ predictions at high temperatures $\beta = 0$.
At the isotropic point, i.e., for the anisotropy $\Delta=1$ we did not
observe evidence of diffusive dynamics at the considered times and momenta.
But even for $\Delta=1$ we found at large momenta a remarkably good agreement
with numerical results for quantum $S=1/2$ density autocorrelations at $\beta
= 0$. On that account we finally argued that our findings for the classical
dynamics may also have impact on the quantum dynamics for $\Delta=1$.
While it is questionable if the present classical approach to transport
is indeed appropriate to simulate possibly pure and strongly $S$-dependent
quantum effects at low temperatures, the possibility of a classical simulation
at not too low temperatures appears to be promising and may be explored
further.

\acknowledgments
The author sincerely thanks W.~Brenig, P.~Prelov\v{s}ek, J.~Herbrych,
T.~Prosen, and M.~{\v{Z}nidari\v{c} for interesting and fruitful
discussions.

\end{document}